\documentclass[aps,prb,reprint]{revtex4-1}
\usepackage{graphicx,float,amssymb,subfigure,dcolumn,epstopdf}
\usepackage{multirow}
\usepackage{bm}
\begin{document}
\title{Dependence of the Inverse Spin Hall Effect in Sr(Nb$_x$Ti$_{1-x}$)O$_3$ on the Nb concentration}
\author{Wanli Zhang, Xiaoyu Zhang, Bin Peng, and Wenxu Zhang\footnote{Corresponding author. E-mail address: xwzhang@uestc.edu.cn}}
\affiliation{State Key Laboratory of Electronic Thin Films and Integrated Devices,
University of Electronic Science and Technology of China, Chengdu, 610054, P. R. China}

\date{\today}
\begin{abstract}
We measured the spin rectify effect and the inverse spin Hall effect in Nb-doped SrTiO$_3$ by injecting the spins from ferromagnetic thin films to SrTiO$_3$ using spin pumping. It was shown that the spin injection is increased when the doping level is increased. However, the spin Hall angle decreases when Nb occupies more than $\sim$2\% of the Ti sites, which is due to that beyond this concentration, the electron contributed the spin Hall effect is from the $d-$orbitals of  Nb instead those from the Ti. Our work points to the importance of orbital occupations in the (inverse) spin Hall effect. We may explore controllable spin and charge interconversion in oxide spintronics.     
\end{abstract}
\maketitle
\section{Introduction}
Manipulation interconversion between electron and spin signals are the essential topic of modern spintronics. The spin Hall effect\cite{Hirsch1999} and its reciprocal part are the two effects where the spin propagation leads to electron flows in certain directions, namely electronic current or vice versa.  The effects were widely explored in heavy metals due to the strong spin orbital coupling (SOC) in them. The spin dependent scattering of electrons are due to the crystal potentials, which is exhibited by the splitting of the energy bands due to the SOC. Impurities are additional scattering centers which were originally proposed to be the source of the spin Hall effect based on the Mott's skew scattering picture. The impurities can also change the resistivity of a material, so that the spin-to-charge conversion efficient can be tuned. 
Systematic study of the spin Hall effect in Ta \cite{Sagasta2018} shows that the spin Hall angle can be tuned by the resistivity of Ta when it is dominant by the intrinsic mechanism. This enhancement of the spin Hall conductivity(SHC) due to the reduction of the longitudinal conductivity was proposed by Sui \emph{et al.} from \emph{ab initio} calculations when $\beta-$tungsten was taken as an example\cite{Sui2017}. Substitution of W with Ta can tune the Fermi level to the gap induced by SOC and the contributions of Berry curvature are enhanced compared with the case that it is canceled if the Fermi level is positioned at one of the SOC-split bands. As we can see, that the extrinsic effects are mainly focused in metals doped by heavy metals or alloys of heavy metals. Giant extrinsic spin Hall effect due to rare-earth impurities was expected because of the strong spin orbital interactions and proper orbital degree of freedom\cite{Tanaka2009}. Lateral spin valve structures are used to obtain the extrinsic spin Hall effects of several doping nonmagnetic metals, Cu or Ag with impurities having a large spin-orbit coupling, Bi or Pb. The spin Hall resistivity is found to be linear with the electric resistivity\cite{Niimi2014}. The spin diffusion length is found to exponentially decrease with the impurity concentration where the resistivity increases due to the impurity scattering probability\cite{Niimi2015}. In addition to the doped metals, conducting 5d oxides shows intrinsic SHE are also studied by Fujiwara \emph{et al.}\cite{Fujiwara2013}, which overcomes the limitations of resistivity encountered in noble metals and Cu-based alloys and shows a very larger spin Hall resistivity $\rho_{SH}$ $\sim$ 38 $\mu\Omega\cdot$cm at the room temperature. 
\par Doping is the key to control the electron transportation in semiconductors. It may as well serve to control the spin-to-charge conversion efficiency, so that spintronics based on semiconductors can be fertilized.  The spin current was able to be generated and transported in $p-$Si at room temperature\cite{Shikoh2013}. The spin diffusion length was estimated to be about 150 nm. The spin Hall effect in a doped semiconductor was calculated by diagrammatic perturbation theory\cite{Tse2006} as well as by solving analytically the kinetic equation including skew scattering at impurities\cite{Engel2005}. The theory of extrinsic spin Hall conductivity in $n-$Doped GaAs was developed by Engel et al. \cite{Engel2005} by treating the SOC as perturbations within the Boltzmann theory of transportation. The spin current was found to be reasonable agreement with experiments by Kato \emph{et al.}\cite{Kato2004} when the parameters are properly taken from experiments. 
\par There are rich properties found in SrTiO$_3$ due to its multi-orbital and the correlated $d-$electrons\cite{Pai2018}. It was reported that Nb-doped SrTiO$_3$ is a good candidate for $n$-type thermoelectric oxide usable at high temperatures\cite{Ohta2005}. The carrier effective mass ($m^\ast$) is about 7 times larger than that of the free electrons which comes from the $d-$orbitals. Large linear magnetoresistance (MR) was observed due to the combination of the interplay between the large classical MR due to high carrier mobility and the electronic localization effect due to strong SOC\cite{Jin2016}. The Nb concentration dependent SOC was further investigated\cite{Cho2018}, which shows a proportionality. It was proposed to be a candidates for channel materials in spintronic devices. 
\par In this work, we measured the ISHE voltage in the Nb-doped STO by spin pumping. DC voltages were obtained at the different dopant concentrations. We found that the spin conversion efficiency decreases upon Nb concentration, which is due mainly to the different orbital occupations when the concentration are increased.



\section{Sample Preparation and measurements}
\begin{figure}
  \centering
  \includegraphics[width=0.5\textwidth]{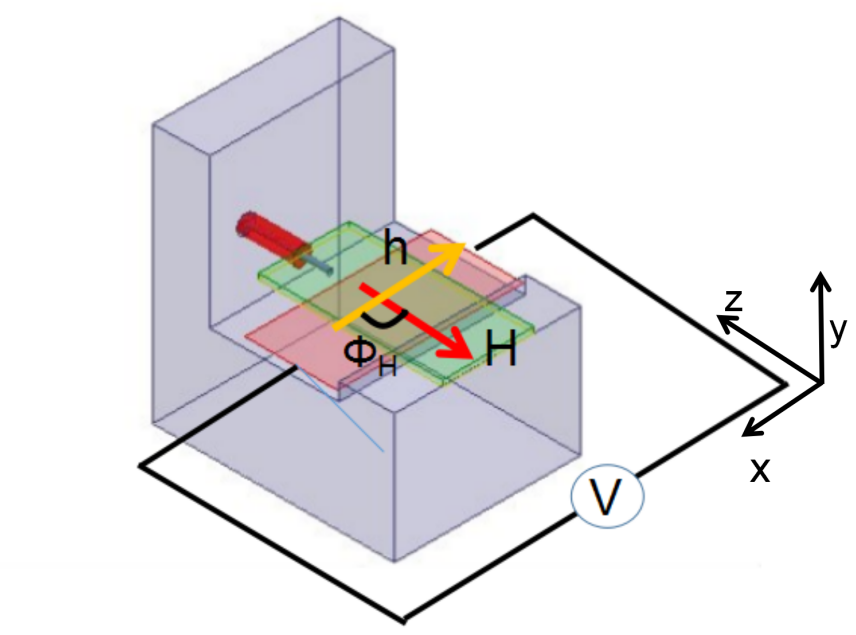}
  \caption{The schematics of the microstrip fixture used in the experiment. The angle between the $x$-axis and the static magnetic field $H$ is $\Phi_H$.}\label{fixture}
\end{figure}

The Nb-doped crystal Sr(Nb$_x$Ti$_{1-x}$)O$_3$ with $x=$ 0.01, 0.015, 0.028, 0.05, 0.1, 0.15, 0.2, and 0.25 are commercially available from the SurfaceNet GmbH. The dimension of the substrate is 10 mm$\times$5 mm$\times$0.5 mm. We grew thin permalloy(Ni$_{80}$Fe$_{20}$,Py) films onto the (001) surfaces of the STO with the film thickness of 20 nm by AC magnetron sputtering. The properties of the ferromagnetic thin film under the RF field was measured by a shorted microstrip fixture which works in a broadband microwave frequency from 2 GHz up to 5 GHz as in our previous works\cite{Zhang2016}. The sample was placed between the center conductor and the ground which is schematically shown in Fig.\ref{fixture}. A static magnetic field $H$ is applied with an angle $\Phi_H$ in the direction of the short side of the sample. The microwave magnetic field is assumed along the long side of our samples in this setup. When the two fields fulfill the ferromagnetic resonance(FMR) condition\cite{Ando2011}, spin currents will be pumped from the magnetic layer to the nonmagnetic SrTiO$_3$, in which the spin currents will be converted into a charge current due to the spin orbit coupling. The so-called Inverse Spin Hall (ISHE) voltage can be measured along the long side of the sample.

\par When the shorted microstrip is used, a microwave magnetic field and a microwave electric field are generated. Due to the AMR effect in the magnetic metallic films, the resistance is alternative which leads to the rectifying effect (SRE) in the metallic magnetic layer. The DC voltage measured then is the ISHE voltage, when exists, mixed with the SRE voltage. In principle this mixture of the signal is always present once a conducting magnetic film is used and placed at the place where the alternating electric field is nonzero.  Based on the fact that the SRE is independent of the spin injection direction, while the ISHE is an odd function of it, if the spin injection direction reverts, the ISHE voltage changes sign and the SRE voltage remains. A method was proposed to separate inverse spin Hall voltage and spin rectification voltage by inverting spin injection direction \cite{Zhang2016} and used in this study.

\section{Results and discussions}

\subsection{FMR and the Magnetic Properties}
\begin{figure}
  \centering
  \includegraphics[width=0.5\textwidth]{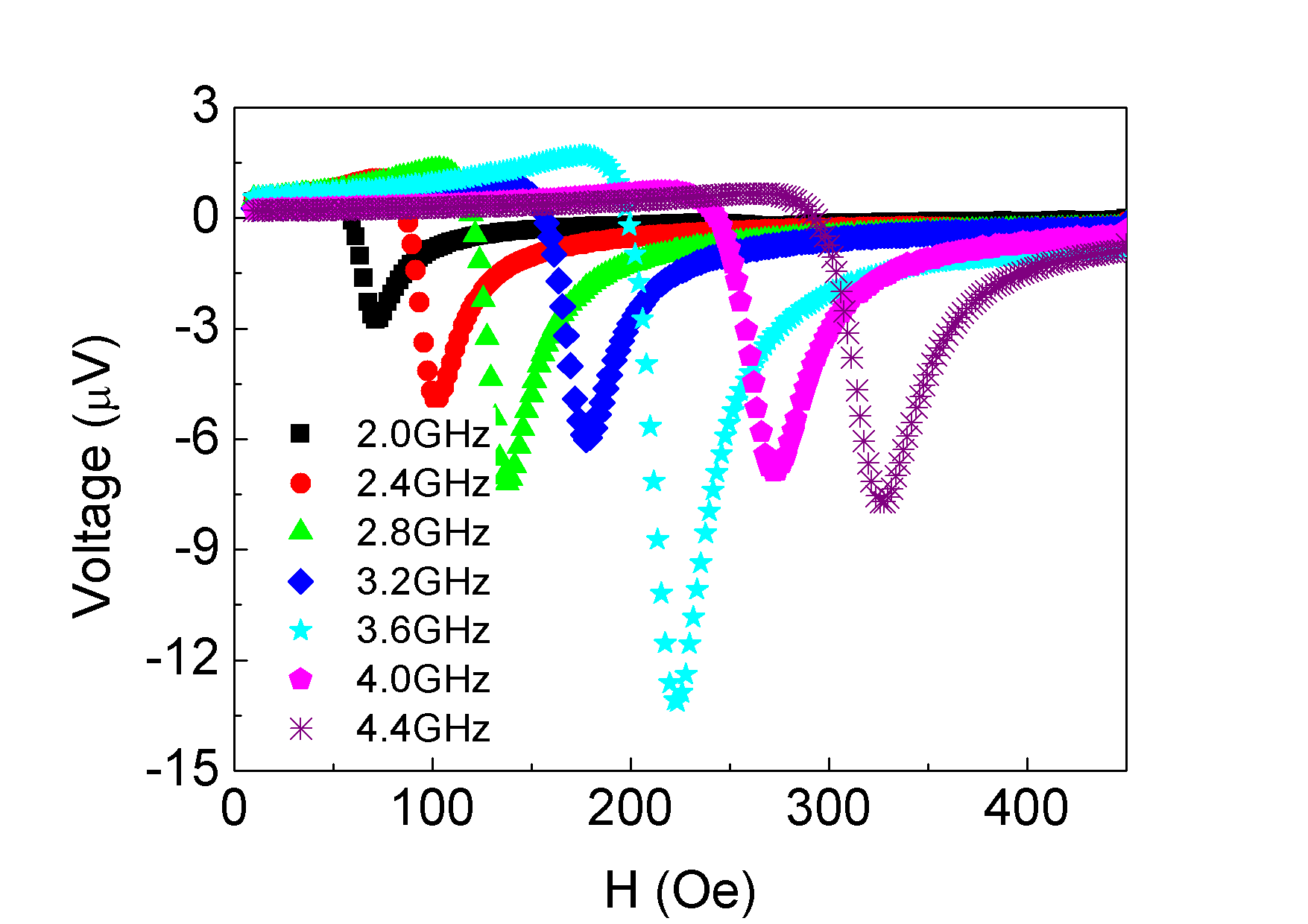}
  \caption{DC Voltage of Py/SrTiO$_3$ measured at the frequencies from 2.0 GHz up to 4.4 GHz with the step of 0.4 GHz. This is the manifest of the SRE effect.}\label{fig:v-h}
\end{figure}

In order to characterize the magnetic properties of the Py thin films, the samples are put into the fixture. DC voltages are measured around the FMR with the microwave frequency varies from 2 GHz up to 4.4 GHz with the step of 0.4 GHz. At each of the different frequencies, a static magnetic field $H$ is varied from 0 to 450 Oe. The voltages obtained at the varied fields and  frequency are shown in Fig.\ref{fig:v-h}.

\begin{figure}
  \centering
  \includegraphics[width=0.4\textwidth]{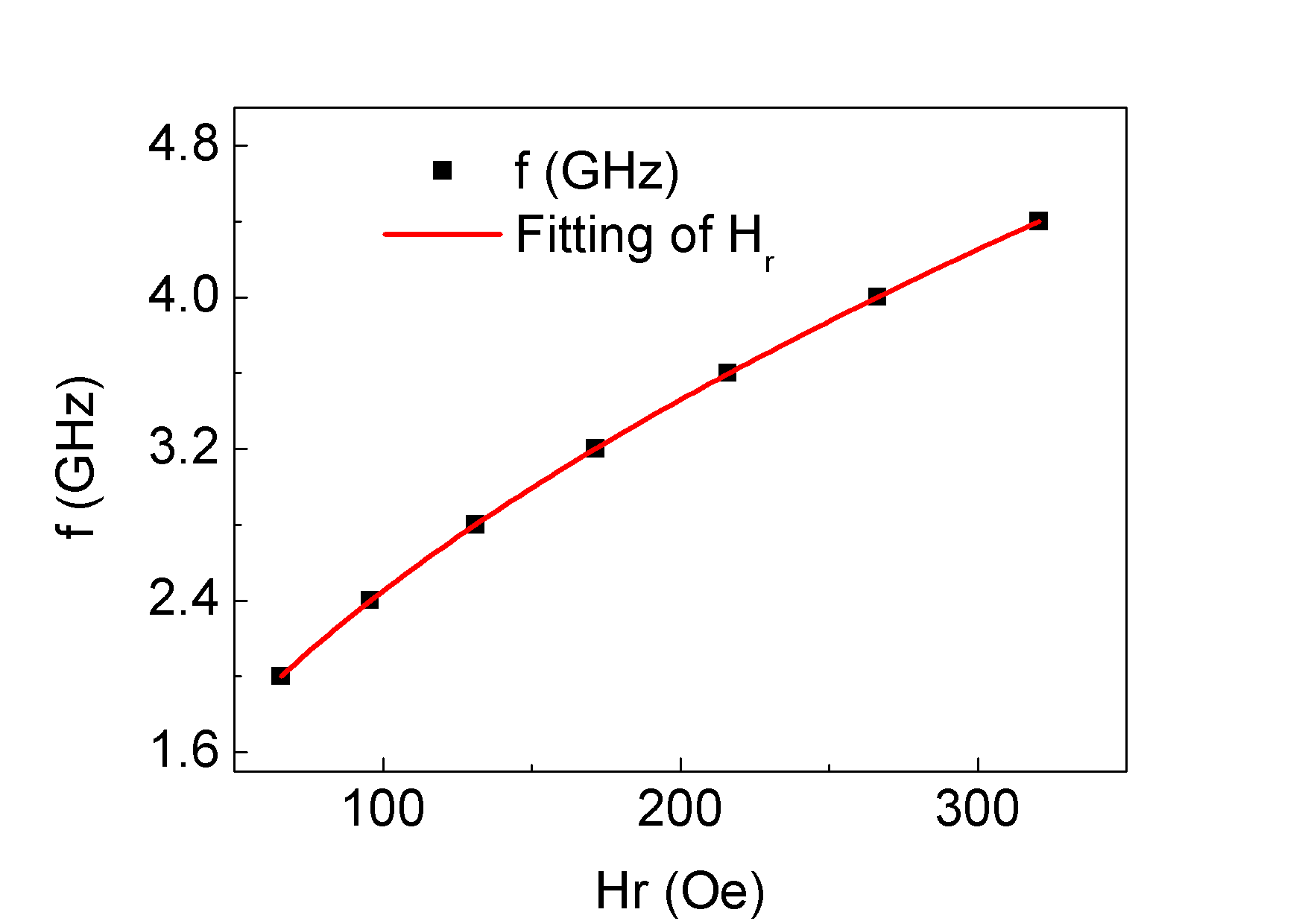}
%
    
  \caption{Magnetic film characterized at FMR: the frequency $f$ varies with resonance field $H_r$ in Py/STO.}\label{fig:kittel}
\end{figure}

As a result of the frequency-dependent microwave propagation and losses within the setup, the intensity of the signal varies when the frequency changes. However, the magnitude of the signal does not play a role in the characterization of the samples. 
The voltage can be obtained by combination of symmetric and antisymmetric Lorenzian lines as in Equ.(\ref{equ:vdc1}) and (\ref{equ:vdc2}): 
\begin{eqnarray}
V_{dc} &=& A_L\cdot L+A_D\cdot D, \label{equ:vdc1}\\ 
L =\frac{\Delta H^2}{4(H-H_r\ )^2+\Delta H^2 } &\text{ and }&
D =\frac{2\Delta H(H-H_r\ )}{4(H-H_r\ )^2+\Delta H^2}, \label{equ:vdc2}
\end{eqnarray}
where $V_{dc}$ is the DC voltage. $A_L$ and $A_D$ are the intensity of the symmetrical and antisymmetric components, respectively. $H_r$ is the resonant magnetic field, and $\Delta H$ is linewidth of the FMR. Thus, by fitting the data obtained in Fig. \ref{fig:v-h}, the frequency dependence of FMR linewidth ($\Delta H$) of the samples, and the frequency dependence of the resonance magnetic field ($H_r$) of the samples can be obtained as shown in Fig. \ref{fig:kittel}. 

According to the Kittel's model for FMR, the resonant magnetic field ($H_r$) and the frequency ($f$) satisfy the following equation:
\begin{equation}
f=\frac{\gamma}{2\pi}\sqrt{\left(H_r+H_k\right)\left(H_r+H_k+4\pi M_s\right)}. \label{equ:k2}
\end{equation}
where $\gamma = 2.21\times 10^5$ m/(A$\cdot$s) is the gyromagnetic ratio, $H_k$ is the magnetic anisotropy field, and  $4\pi M_s$ is the saturation magnetization. 

\par
\begin{figure}
	\centering
	\includegraphics[width=0.5\textwidth]{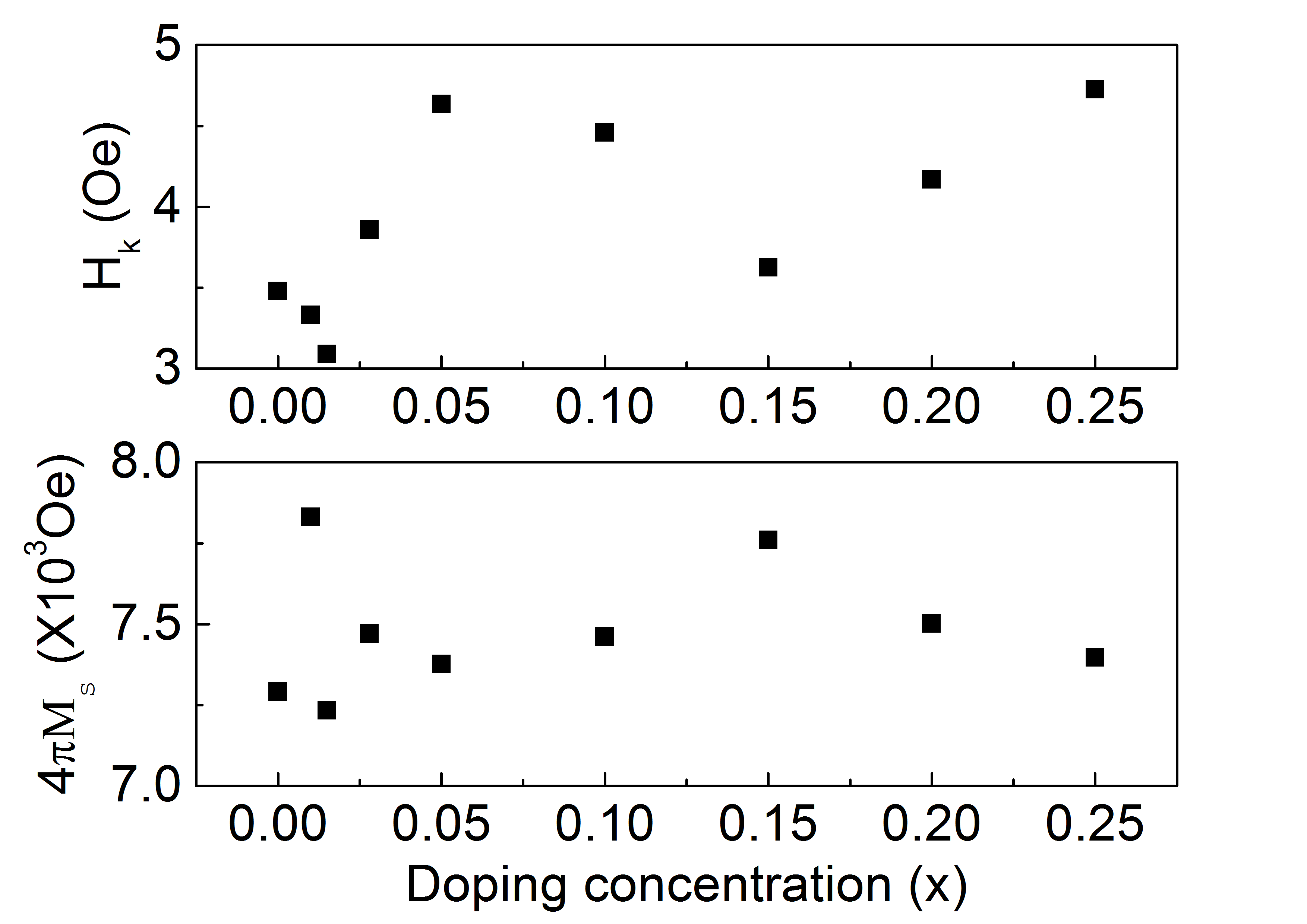}
	\caption{The effective saturation magnetization $4\pi M_s$ and the anisotropy constant $(H_k)$ of the magnetic films deposited on the STO with different doping concentration.}\label{fig:ms_hk}
\end{figure}

 By the fitting, we can obtain the magnetic anisotropic field and the saturation magnetization of the Py films deposited on the substrates with different doping concentrations as shown in Fig. \ref{fig:ms_hk}. The anisotropic field varies between 3 and 5 Oe while the saturation magnetization is between 7200 and 7800 Oe. As we prepared the films under the same condition, the two magnetic parameters are the same between the samples within the accuracy of the experiments. The dopant concentration of the substrate does not influence the magnetic properties of the thin films.
\par The damping coefficient $\alpha$ of the different samples can be obtained by fitting the relationship between the linewidth ($\Delta H$) and the frequency ($f$) to the following linear expression:
\begin{equation}
\Delta H=\Delta H_0+\frac {4\pi\alpha f}{\sqrt3\ |\gamma|\ }, \label{equ:k1} 
\end{equation}
where $\Delta H_0$ is the intrinsic linewidth, and $\alpha$ is the damping coefficient. 

\begin{figure}
	\centering
	\includegraphics[width=0.5\textwidth]{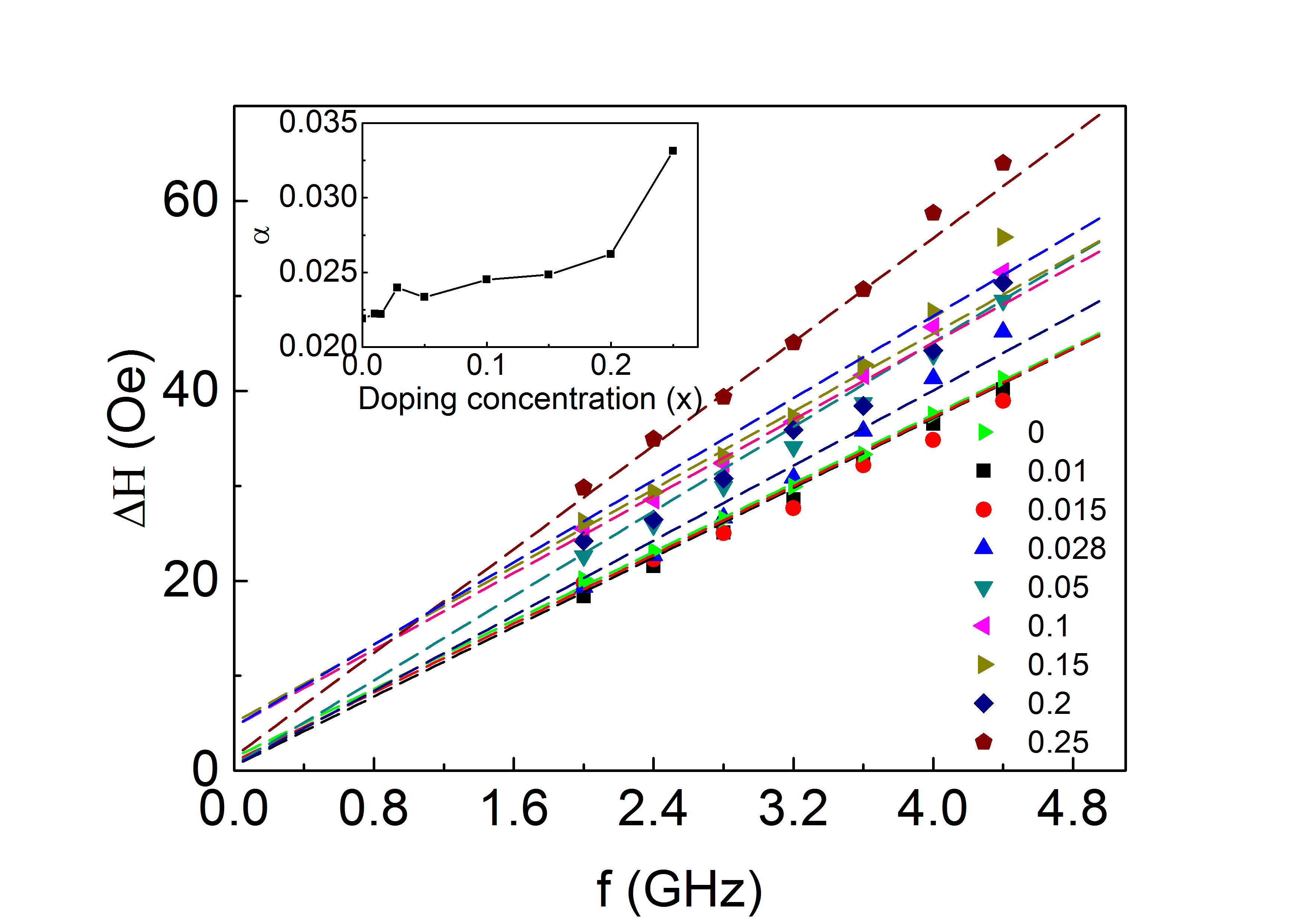}
	\caption{The relationship between the linewidth $\Delta H$ and the frequency $f$. The damping coefficient $\alpha$ at different doping levels (x) obtained by fitting is shown in the inset. }\label{fig:deltaH}
\end{figure}

\par The data and the fitted lines are shown in Fig. \ref{fig:deltaH}. It can be observed that the data can be fitted well with the linear equation. The intrinsic linewidth clusters to below 4 Oe indicating the magnetization damping due to the intrinsic defects are almost the same and very low. The damping coefficient $\alpha$ is clearly the function of doping concentration. It increases from 0.022 to 0.033 when the Nb concentration $x$ increases from 0.00 to 0.25. The increment of the damping coefficient is the signature of the increment of the spin angular momentum losses. This is mainly due to the enhanced conductivity of the STO substrate with the increased doping levels. Within the diffusion theory, the spin relaxation time is proportional to the electron relaxation time. Within this theory, the spin diffuses into the STO more easily when the electron conductivity of the sample is higher.  
%

\subsection{Inverse spin Hall voltage and the Spin Hall Angle}
\begin{figure}
  \centering
  \includegraphics[width=0.5\textwidth]{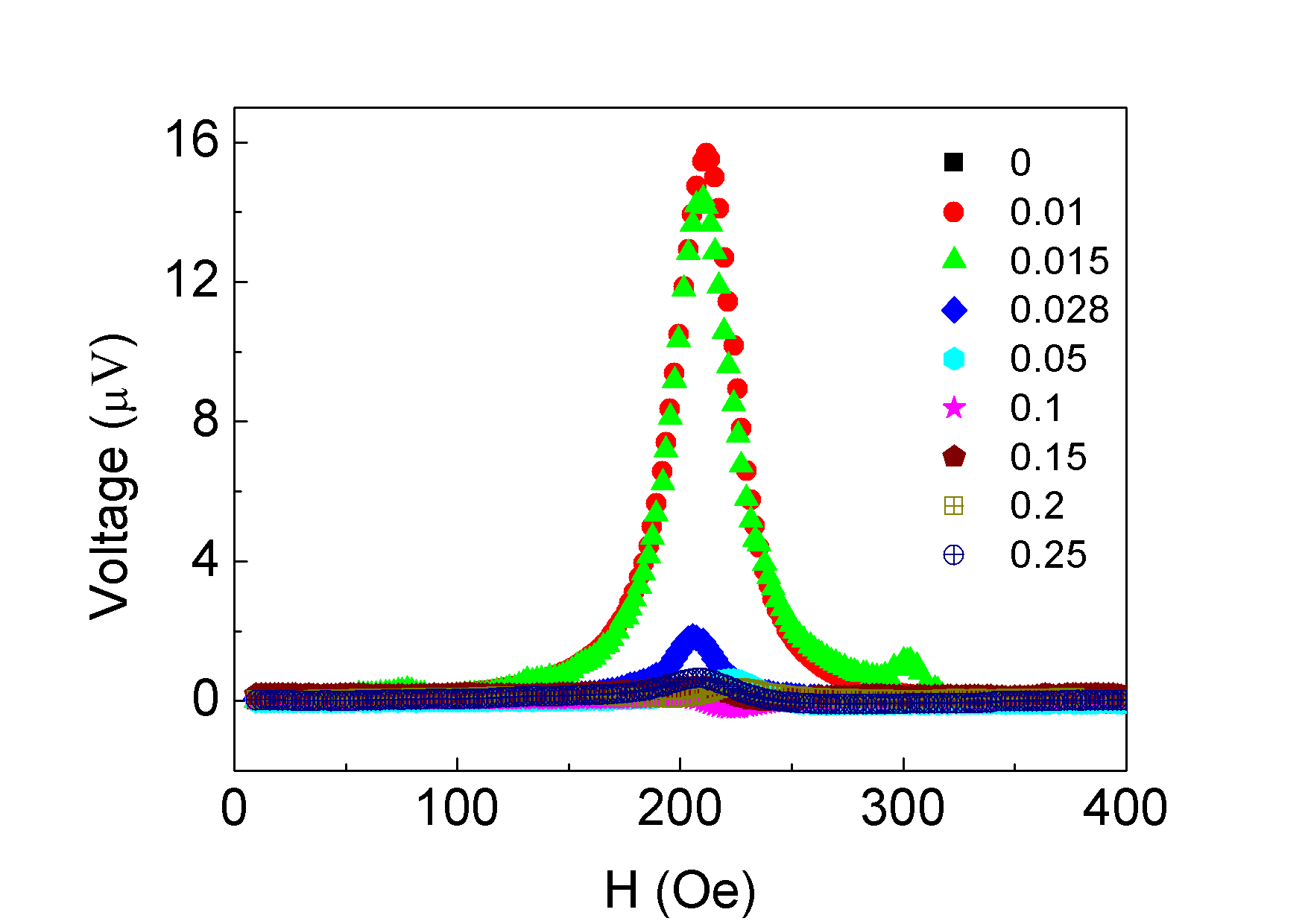}
  \caption{(color online) The ISHE voltage with the different doping concentrations measured under the external magnetic field at 3.6 GHz.}\label{fig:ishe}
\end{figure}

The inverse spin Hall voltages of the samples with different doping concentrations obtained from our measurement with the microwave power $P = 600$ mW and frequency $f = 3.6$ GHz are shown in Fig.\ref{fig:ishe}.

It can be found from Fig.\ref{fig:ishe} that at the doping levels of $x$ = 0.01, 0.015 and 0.028 the Py/Sr(Nb$_x$Ti$_{1-x}$)O$_3$ samples have obvious inverse spin Hall voltage, while in the other samples the inverse spin Hall signals are negligible. The intrinsic STO  ($x=0$) is insulating, so that the spins generated in Py cannot diffuse through the interface of the Py/STO. As the result, there is no ISHE voltage detectable as shown by the black filled squares in the figure. For the other samples with higher Nb doping concentration ($x \geq 0.028$), the ISHE voltage is however diminishing. 
\par When the spin current is injected into the doped STO, the inverse spin Hall voltage ($V_{ISHE}$) measured in the samples can be expressed\cite{Ando2011} as Equ. (\ref{equ:vishe}):
\begin{equation}
V_{ISHE}=\frac{l\theta\lambda_N\tanh{\left(d_N/2\lambda_N\right)}}{d_N\sigma_N+d_F\sigma_F}\left(\frac{2e}{\hbar}\right)j_s^0.\label{equ:vishe}
\end{equation}
with the spin current $j_s^0$:
\begin{equation}
j_s^0=\frac{g\gamma^2{h_{RF}}^2\hbar\left[4\pi M_s\gamma+\sqrt{\left(4\pi M_s\right)^2\gamma^2+4\omega^2}\right]}{8\pi\alpha^2\left[\left(4\pi M_s\right)^2\gamma^2+4\omega^2\right]},\label{equ:js}
\end{equation}
where $l$ is the length of the sample. $\theta$ and $\lambda_N$ are the spin-Hall angle and the spin diffusion length of the Nb doped STO. $\sigma_{N(F)}$ is the conductivity of the nonmagnetic(ferromagnetic) layer. $d_N(F)$ is the thickness of them. $e$ is the electron charge, $\hbar$ is the Dirac constant, and $g$ is the real part of the spin mixing conductance. $h_{RF}$ is the microwave magnetic field magnitude with angular frequency of $\omega$.
\par
\begin{figure}
  \centering
  \includegraphics[width=0.5\textwidth]{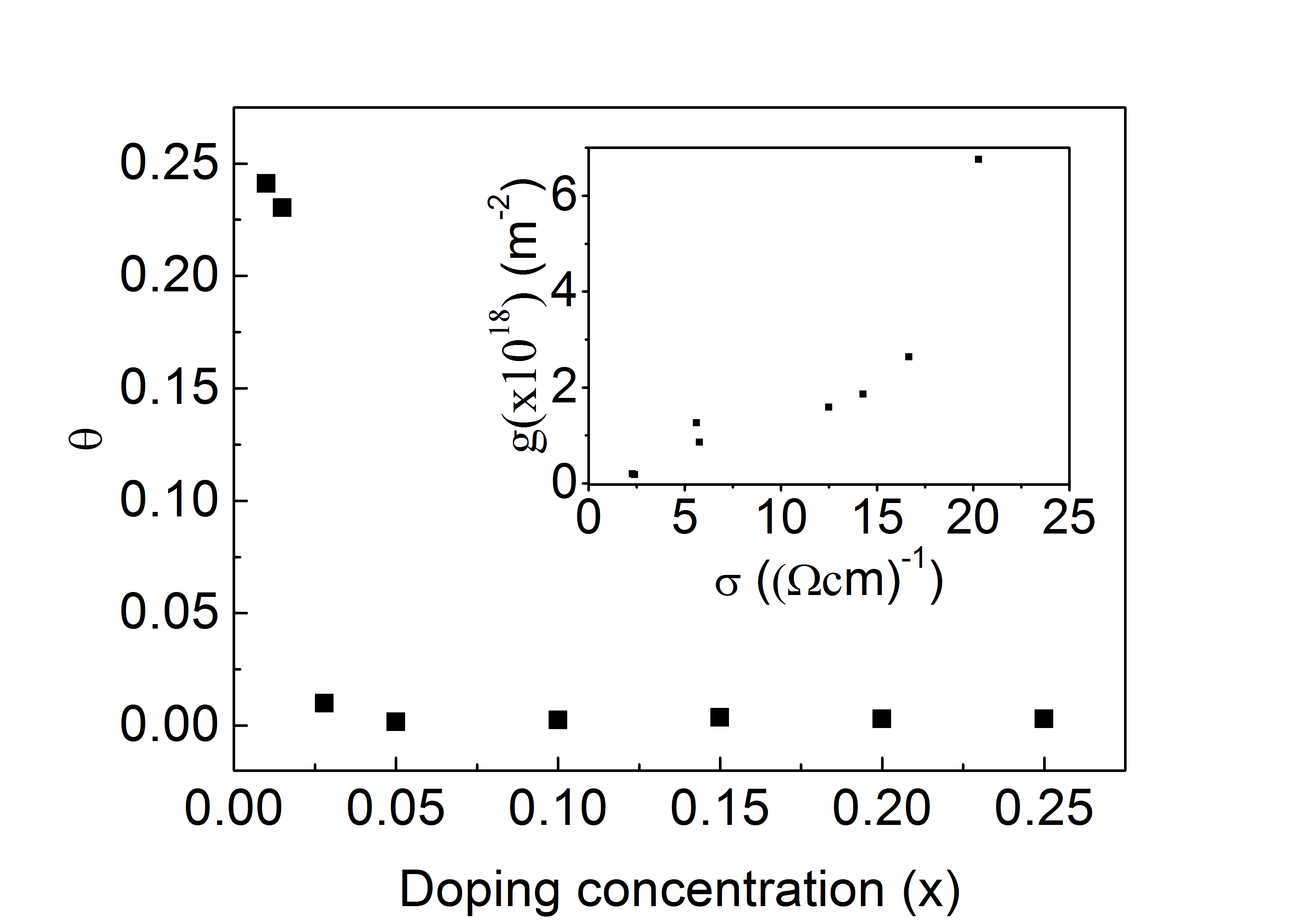}
  \caption{The spin Hall angle ($\theta$) vs. the doping concentrations (x). The spin mixing conductance $g$ with respect to the electron conductivity $\sigma$ is shown in the inset.}\label{fig:hallanlge}
\end{figure}

 The spin mixing conductance $g$ is calculated by
\begin{equation}
g=\frac{2\sqrt3\pi M_s\gamma d_F}{g_L\mu_B\omega}\left(\Delta H_{F/N}-\Delta H_F\right), \label{equ:g}
\end{equation}
where $g_L$ is the Land\'{e} factor, $\mu_B$ is the Bohr magneton, and $\Delta H$ is the linewidth at ferromagnetic resonance. The parameters used in our calculation are $\gamma=2.21\times10^5$ m/(A$\cdot$s), $\omega=2.262\times10^{10}$ rad/s, and $d_F=20$ nm.
In order to estimate the microwave magnetic field magnitude, we calculated the power absorption of the samples at the FMR. When the microwave power density is not very large, the relationship between the power absorption ($\Delta P$) in the Py films and the microwave magnetic field ($h_{RF}$) can be expressed as\cite{Iguchi2012}:
\begin{equation}\Delta P=V\frac{\mu_0\gamma4\pi M_s}{4\alpha}h_{RF}^2\left(\frac{\gamma4\pi M_s+\sqrt{\left(\gamma4\pi M_s\right)^2+\left(2\omega\right)^2}}{\sqrt{\left(\gamma4\pi M_s\right)^2+\left(2\omega\right)^2}}\right). \label{equ:power}
\end{equation}
Using Equ. (\ref{equ:vishe}),(\ref{equ:js}), (\ref{equ:g}) and (\ref{equ:power}) with parameter $d_N=0.5$ mm, we obtain the spin-Hall angle ($\theta$) of the samples with different doping concentration as in Fig. \ref{fig:hallanlge}.

We first notice that the spin Hall angle is about 0.22, which is the same order of magnitude as that of the heavy metals, such as Ta $\sim$ 0.1. However, the spin Hall angle decreases with the dopant concentration, although the spin mixing conductance is increased as shown in the inset of the same figure.  This is on the contrary to the case of doping in metals, which means that there are more spins diffuse into the nonmagnetic substrate, the conversion efficiency is largely reduced. The conversion efficiency of the spin into the charge current is complex as exampled in this work. 
\par As widely accepted that the extrinsic mechanism of spin Hall conductivity has two sources, namely the side-jumping (SJ) and skew scattering (SS). The conductivity obtained by the sophisticated Green's function calculations\cite{Tse2006} shows that the SJ contribution to the spin Hall conductivity is 
\begin{equation}
\sigma_{xy}^{SJ}=\frac{e^2\lambda_0^2}{4\hbar}n,
\end{equation}
where $\lambda_0$ is the SOC strength and $n$ is the electron density.
Within the Drude model, the longitudinal electron conductivity can be written as $\sigma_{xx}=ne^2\tau/m$. The spin Hall angle can be written as
\begin{equation}
\theta^{SJ}=\frac{\sigma_{xy}^{SJ}}{\sigma_{xx}}=\frac{m\lambda_0^2}{4\hbar}\frac{1}{\tau}.
\end{equation}
That from the SS reads as
\begin{equation}
\theta^{SS}=-\pi m\lambda_0^2\epsilon_F/3\hbar^2. 
\end{equation}
By this consideration, the spin Hall angle is proportional to $\lambda_0^2$. According to the estimation of Cho \cite{Cho2018}, the SOC field $B_{SO}$ is about 1 T when $x=0.02$, increase with a rate of 0.21 T/at.\% with the dopant concentration. However, according to our results, the spin Hall angle decreases with the doping concentration which points to the conclusion that the effective mass is greatly reduced by doping. According to the work by Cho \emph{et al.} \cite{Cho2018}, when the concentration of Nb larger than 0.02, the Fermi level situates above the orbital of Ti $d_{z^2,x^2-y^2}$ and enters the states contributed by the Nb $d-$electrons. In this case, we see the importance of the individual orbitals which determine the SHE. We also note that in recent work on Ta\cite{Sagasta2018}, the authors show that the enhancement of the spin Hall angle by increase the resistivity of Ta. The same tendency was observed also in CuBi alloys\cite{Niimi2014}. Although our work shows the same tendency, the mechanism is different.
\par We may take some order-of-magnitude estimation of the parameters for the samples with obvious spin Hall effect. The relaxation time $\tau$ in STO is typically $10^{-13}-10^{-12}$, the spin Hall angle of $0.22$, and the effective electron mass $m\sim7.5 m_0$, we get $\lambda_0$ of order $1\times10^{-9}$ m, which is a factor of only about 100 times  enhancement over the vacuum electron Compton wavelength of $3.9\times10^{-11}$ m. This is at the same order of magnitude as that of $p-$doped GaAs\cite{Tse2006}. 

\section{Conclusions}
We have shown in this work, that doping STO with atoms endowed with considerable SOC, the spin current can be converted to the charge current. However, the conversion efficient does not proportional to doping level. In our work, the ISHE can be observed when the doping concentration is below 0.02 \%. When the concentration is beyond that level, the conversion of spin charge to the charge current is negligibly small. Also the spins diffuse more efficiently to the STO layer due to the enhancement of the conductivity, the conversion of the spin current to the charge current is dependent on the orbitals.
\section{Acknowledgments}
Financial support from National Key R\&D Program of China (No.2017YFB0406403) was greatly acknowledged.

\end{document}